\def\slaninafigpath{}
\newcommand{\slaninaepsfigure}[2]{
  \includegraphics{\slaninafigpath#1.eps}
}
\newcommand{\slaninafigureh}[2]{
\begin{figure}[hb]
  \centering
  \vspace*{75mm}
  \includegraphics{\slaninafigpath#1.ps}
  \caption{#2}
  \label{fig:#1}
\end{figure}
}
\newcommand{\slaninafigurehh}[2]{
\begin{figure}[hb]
  \centering
  \vspace*{100mm}
  \includegraphics{\slaninafigpath#1.ps}
  \caption{#2}
  \label{fig:#1}
\end{figure}
}

\documentstyle[pre,aps]{revtex}
\begin{document}
\draft

\twocolumn[\hsize\textwidth\columnwidth\hsize\csname@twocolumnfalse\endcsname

\title{Random networks created by biological evolution} 

\author{Franti\v{s}ek Slanina \cite{f-adr} and Miroslav Kotrla \cite{m-adr}}
\address{Institute of Physics, Academy of Sciences of the Czech Republic,\\
Na Slovance 2, CZ-182~21 Praha 8, Czech Republic}
%\date{\today}

\maketitle
\begin{abstract}
We investigate a model of evolving random network, introduced
by us previously {[}{\it Phys. Rev. Lett.} {\bf 83}, 5587 (1999){]} . The
model is a generalization 
of the Bak-Sneppen model of biological evolution, with the
modification that the underlying network can evolve by adding and
removing sites. The behavior and the averaged properties of the
network depend on the 
parameter $p$, the probability to  
establish link to the newly introduced site. For $p=1$ the system
is self-organized critical, with two distinct power-law regimes with
forward-avalanche exponents  
$\tau=1.98\pm 0.04$ and $\tau^\prime = 1.65\pm 0.05$.
The average size of the network diverge as power-law when $p\to 1$. 
We study various geometrical properties of the network: probability
distribution of sizes and connectivities, size and number of
disconnected clusters and the dependence of mean distance between two
sites on the cluster size. The connection with models of growing
networks with preferential attachment is discussed.
\end{abstract}

\twocolumn]

\section{Introduction}

Irregular networks or random graphs \cite{bollobas_85} composed of units of various kind are very frequent
both in nature and society (which is, however, nothing but a special
segment of nature). Examples range from vulcanized polymers, silica
glasses, force chains in granular materials \cite{ra_je_mo_ro_96},
mesoscopic quantum wires \cite{kott_smi_98} to  
food webs \cite{wil_mar_00},
herding effects in economics \cite{co_bou_97}, world-wide-web links
\cite{bar_alb_99,alb_jeo_bar_00}, 
``small-world'' networks of personal contacts between
humans \cite{wa_stro_98,newman_00} and scientific collaboration
networks \cite{newman_00a}.

Modeling of such networks is not quite easy and analytical results
are relatively rare (examples, without any pretence of completeness,
can be found in \cite{bollobas_85,co_bou_97,new_moo_wat_99,kul_alm_st_99a}). 
Numerical simulations are still one of the
principal tools. However, even in the case when the properties of a
given class of random networks are relatively well established, either
analytically or numerically, as is the case of small-world networks,
the serious question remains, why do these networks occur in
nature. In other words, what are the dynamical processes, which
generate these networks. 

Indeed, one can study, for example, various networks of mutual
dependence of species in a model of co-evolution
\cite{kauffman_90a,ba_sne_93,so_ma_96}, but it is 
difficult to infer from these studies only, which networks are closer
to the reality that the others. In the context of biological evolution models,
there was recently a few attempts to let the networks evolve freely, in
order to check, which types of topologies might correspond to
``attractors'' of 
the process of natural evolution \cite{chri_do_ko_sne_98,ja_kri_98,sla_ko_99,bor_roh_00,bor_sne_98,bor_sne_00}.

The model introduced by us in a preceding Letter \cite{sla_ko_99} is based
on extremal dynamics and basically follows the Bak-Sneppen
model of biological evolution \cite{ba_sne_93}.
Extremal dynamics (ED) models \cite{pa_ma_ba_96} are used in wide area
of problems, ranging 
from growth in disordered medium \cite{sneppen_92}, dislocation
movement \cite{za_92}, friction \cite{slanina_98} to biological
evolution \cite{ba_sne_93}. Among them, the Bak-Sneppen (BS) model
plays the role of a testing ground for various
analytical as well as numerical approaches (see for example
\cite{pa_ma_ba_96,grassberger_95,bo_ja_we_95,pismak_95,ma_lor_ma_97,delos_va_ve_97,va_au_95}). 

The idea of ED is the following.
The 
dynamical system in question is composed of a large number of
simple units, connected in a network. Each site of the network hosts
one unit. The state of each unit is described
by a single dynamical variable $b$, called barrier. In each step,
the unit with minimum $b$ is mutated by updating the barrier.
The effect of the mutation on the
environment is taken into account by changing $b$ also at all
sites connected to the minimum site by a network link. 
Because a perturbation can propagate
through the links, we should expect that the topology of the network
can affect substantially the ED evolution.

General feature of
ED models is the avalanche dynamics. The forward $\lambda$-avalanches
are defined as follows \cite{pa_ma_ba_96}. For fixed $\lambda$ we define
active sites as those having barrier $b < \lambda$. Appearance
of one active site can lead to avalanche-like proliferation of active
sites in successive time steps. The avalanche stops, when all active
sites disappear  
again. Generically, there is a value of $\lambda$, for which the probability
distribution of  avalanche sizes obeys a power law without any
parameter tuning, so  that the ED models are classified as a subgroup
of self-organized critical models \cite{ba_ta_wi_87}. (This, of course,
can hold only for networks of unlimited size.) The set of exponents
describing the critical behavior determines the dynamical universality
class the model belongs to.

It was found that 
the universality class
depends on the topology of the network. Usually, regular hyper-cubic
networks \cite{pa_ma_ba_96} 
or Cayley trees 
\cite{va_au_95} are 
investigated. For random neighbor networks, mean-field solution
was found to be exact
\cite{debo_de_fly_ja_we_94,bo_ja_we_95}. Also the tree models  \cite{va_au_95}
were found to belong to the mean-field universality class.
A one-dimensional model in which the links were wired randomly with
probability decaying as a power $\mu$ of the distance was introduced
\cite{ca_delos_val_ve_99,gle_ta_can_99}. It was found that the values of critical exponents depend continuously on $\mu$.
The BS model on a small-world network was also studied \cite{kul_alm_st_99}.

Recently, BS model on random networks, produced by bond percolation on
fully connected lattice, was studied \cite{chri_do_ko_sne_98}. Two
universality classes were found. Above the percolation threshold, the
system belongs to the mean-field
universality class, while exactly at the
percolation threshold, the avalanche exponent is different.
 A dynamics changing the topology in order to drive the
network to critical connectivity was suggested. 

There are also several recent results for random networks
produced by different kind of dynamics than ED, especially for the
threshold networks \cite{bor_roh_00} and Boolean networks
\cite{bor_sne_98,bor_sne_00}. 

The geometry of the world-wide web was intensively studied very
recently. It was found experimentally that the network exhibits   
scale-free characteristics, measured by the power-law distribution of
connectivities of the sites \cite{bar_alb_99,bar_alb_jeo_00}.
Similar power-law behavior was observed also in the actor
collaboration graph and in the power grids \cite{bar_alb_99}. A model
was suggested \cite{bar_alb_99} to explain this behavior, whose two
main ingredients 
are continual growth and preferential attachment of new links, where
sites with higher connectivity having higher probability to receive
additional links. The latter feature resembles the behavior of
additive-multiplicative random processes, which are well known
to produce power-law distributions \cite{so_co_97,sornette_98c}.

The model introduced in \cite{bar_alb_99} is exactly soluble
\cite{dor_men_sam_00}. Variants including aging of sites
\cite{dor_men_00,kra_red_ley_00},  decaying and rewiring links
\cite{alb_bar_00,dor_men_00a} were also studied.
The preferential attachment rule, which apparently requires
unrealistic knowledge of connectivities of the whole network before a
single new link is established, was justified in a very recent work
\cite{vazquez_00}, where higher probability of attachment at highly
connected sites results from local search by walking on the network.

In the preceding Letter \cite{sla_ko_99} we concentrated on the
self-organized critical behavior and extinction dynamics of a model in
which the network changes dynamically by adding and removing sites. It
was shown that the extinction exponent is larger than the upper bound
for the BS model (given by the mean-field value) and is closer to the
experimentally found value than any previous version of the BS
model. In the present work we introduce in Sec. \ref{sec:model} 
a generalized version of the
model defined in \cite{sla_ko_99} and further investigate the self-organized
critical behavior in Sec. \ref{sec:soc}. 
However, our main concern will be about the
geometric properties of the network, produced during the
dynamics. These results are presented in Sec. \ref{sec:geometry}. 
Section \ref{sec:conclusions} makes
conclusions from the results obtained.

\section{Evolution model on evolving network}
\label{sec:model}

We consider a system composed of varying number $n_{\rm u}$ of
units connected in a network, subject to extremal dynamics. 
Each unit bears a dynamical variable
$b$. In the context of biological evolution 
these units are species and $b$ represent the barrier against mutations.  
For the main novelty of our model consists in adding (speciation) and
removing (extiction) units, let us first define the rules for
extinction and speciation. The rules determining which of the existing
units will undergo speciation or extinction will be specified
afterwards.

{\it (i) } If a unit is chosen for extinction, it is completely
removed from the network
without 
any substitution and all links it has, are broken.

{\it (ii) }
If a unit is chosen for speciation, it acts as a ``mother'' giving
birth to a new, ``daughter'' unit. A new unit is added into the system
and the links are established between the new  unit and the 
neighbors of the ``mother'' unit: each link of the ``mother'' unit is
inherited with the probability $p$ by the ``daughter'' unit. 
This rule reflects the fact that the new unit is to a certain extent 
a copy of the 
original, so the relations to the environment will be initially
similar to the ones the old unit has. 
Moreover,
if a unit which speciates has only one neighbor, a link between
``mother'' and ``daughter'' is also established.

The extremal dynamics rule for our model is the following.

{\it (iii) } In each step, the unit
with minimum $b$ is found and mutated.
 The barrier of the mutated unit is replaced by a new random
value $b^\prime$ taken from uniform distribution on the interval
$(0,1)$.
Also the barriers of all its neighbors are replaced
by new random numbers from the same distribution.

The rules determining whether a unit is chosen for extiction or
speciation are the following.
 
{\it (iv) }
If the newly assigned barrier of the mutated unit $b^\prime$ is larger
than new barriers of all its neighbors, 
the unit is chosen for speciation. If $b^\prime$ 
is lower than barriers of all neighbors, the 
unit is chosen for extinction. In other cases neither extinction nor
speciation occcurs. As a boundary condition, we use the
following exception: if the network
consists of a single isolated unit only, it is always chosen for speciation.

{\it (v) } If a unit is chosen for extinction, all its neighbors which
are not connected to any other unit also chosen for extinction. 
We call this kind of
extinctions singular extinctions.

The rule {\it (iv)} is motivated by the following considerations. We assume,
that well-adapted units 
proliferate more rapidly and chance for speciation is bigger. However,
if the local biodiversity, measured by connectivity of the unit, is
bigger, there are fewer empty ecological niches and the probability of 
speciation is lower. On the other hand,
poorly adapted units are more vulnerable to extinction, but at the
same time larger biodiversity (larger connectivity) may favor the survival.
Our rule corresponds well to these assumptions: speciation occurs
preferably at units with high barrier and surrounded by fewer
neighbors, 
extinction is more frequent at units with lower barriers and lower
connectivity. Moreover, we suppose that a unit completely isolated
from the rest of the ecosystem has very low chance to survive. This
leads to the rule {\it (v)}.

From the rule {\it (iv)} alone follows equal probability of
adding and removing a unit, because the new random barriers $b$ are
taken from uniform distribution. At the same time the rule {\it (v)}
enhances the probability of the removal.
Thus, the probability of speciation is slightly lower than
the probability of extinction. 
\begin{figure}[hb]
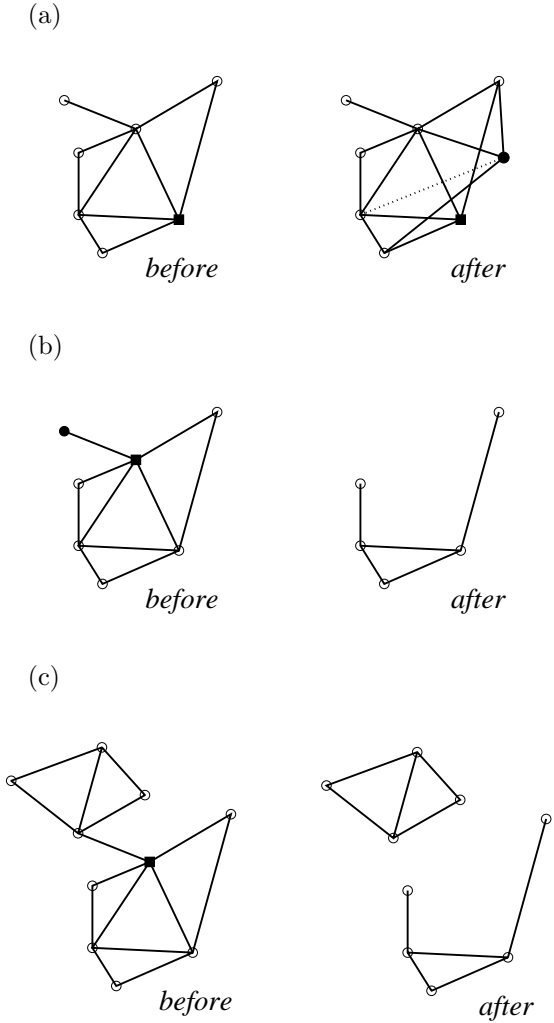

\hspace*{10mm}(a)
  \vspace*{40mm}

\slaninaepsfigure{speciation}{50}

%  \special{psfile=speciation.eps
%        hscale=40 vscale=40
%	voffset=20 hoffset=50 
%        angle=0}

\hspace*{10mm}(b)
\vspace*{40mm}

\slaninaepsfigure{extinction}{50}

%  \special{psfile=extinction.eps
%        hscale=40 vscale=40
%	voffset=20 hoffset=50 
%        angle=0}

\hspace*{10mm}(c)
\vspace*{50mm}

\slaninaepsfigure{extinction-break}{30}

%  \special{psfile=extinction-break.eps
%        hscale=40 vscale=40
%	voffset=20 hoffset=30 
%        angle=0}

  \caption{Schematic illustration of the dynamical rules of the
        model. Speciation is shown in (a), where full square represents
        the extremal unit, which speciates, full circle the new,
        daughter unit, and open circles other units, not affected by
        the speciation event. The dotted link intends to illustrate that
        for $p<1$ some of the mother's links may not be inherited by
        the daughter. Extinction is shown in (b), where the extremal
        unit, which is removed, is indicated by full square. The unit
        denoted by full circle is the neighbor removed by the singular
        extinction. In (c) an example of an extinction event is shown,
        which leads to the 
        splitting of the network into disconnected clusters.}
  \label{fig:illustration}
\end{figure}
The degree of disequilibrium between the two
depends on the topology of the network at the moment and can be
quantified by the frequency of singular extinctions. The number of
units $n_{\rm u}$ perform a biased random walk with reflecting
boundary at $n_{\rm u} = 1$. The bias towards small values is not
constant, though, but fluctuates as well. 

The above rules are illustrated by the examples shown in 
Fig. \ref{fig:illustration}. The networks in (a) show the effect of the
speciation: a new site is created and some to the links to the
mother's neighbors are established.  In (b) the extinction is
shown. One of the units is removed also due to a singular extinction
(rule {\it (v)}). In (c) we illustrate the possibility that in the
extinction event the
network can be split into several disconnected clusters.

\section{Self-organized critical behavior}
\label{sec:soc}
\subsection{Crossover scaling}

The model investigated in the preceding Letter \cite{sla_ko_99}
corresponds to the value $p=1$. 
We found that in this case the model is self-organized critical. We
defined newly the mass extinctions, as number of units removed during
an avalanche. The distribution of mass extinctions obeys a power law
with the exponent $\tau_{\rm ext}=2.32\pm 0.05$. 
In this section we present improved analysis of the data for
the self-organized critical behavior.

\slaninafigureh{aval-new-scaled}{
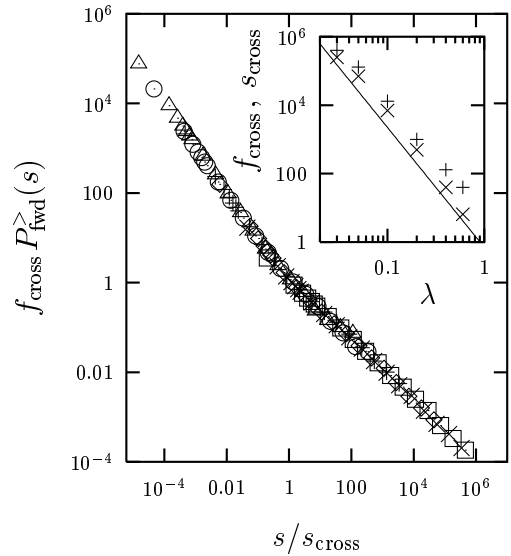
Rescaled distribution of forward avalanches in the case $p=1$, for the values 
$\lambda=0.03$ ($\triangle$), %8
$\lambda=0.05$ ({\Large $\circ$}), %6 
$\lambda=0.1$ ($+$),  %1
$\lambda=0.2$ ($\Diamond$),  %12
$\lambda=0.4$ ($\times$),  %2
$\lambda=0.6$ ($\Box$),  %4  
The superscript $>$ in $P_{\rm fwd}^>(s)$ is to indicate that we count all
avalanches larger 
than $s$.
In the inset we plot the dependence of the scaling parameters $s_{\rm
cross}$ ($+$) and $f_{\rm cross}$ ($\times$) on $\lambda$. The full
line is the power-law $\lambda^{-\sigma^\prime}$ 
with the exponent $\sigma^\prime=3.5$. The number of time steps was
$3\cdot 10^8$ and the data are averaged over 12 independent runs.
}

We
measured the distribution of forward
$\lambda$-avalanches
\cite{pa_ma_ba_96}
and we observed, contrary to the BS model that two power-law regimes
with two different 
exponents occur. The crossover value $s_{\rm cross}$ which separates
the two regimes depends on $\lambda$. We observed that the
distributions for different $\lambda$ collapse onto single curve, if 
plotted against the rescaled avalanche size $s/s_{\rm cross}$,
i. e.
\begin{equation}
P^>_{\rm fwd}(s)\cdot f_{\rm cross} = g(s/s_{\rm cross})
\label{eq:collapse}
\end{equation}
where $g(x)\sim x^{-\tau+1}$ for $x\ll 1$ and $g(x)\sim x^{-\tau^\prime+1}$ 
for $x\gg 1$. The data are plotted in Fig. \ref{fig:aval-new-scaled}.
For the values of the exponents, we found $\tau= 1.98\pm 0.04$ and
$\tau^\prime=1.65\pm 0.05$.

We investigated the dependence of the scaling parameters 
$s_{\rm cross}$ and $f_{\rm cross}$ on $\lambda$ and we found that both of
them behave as a power law with approximately equal exponent, 
$s_{\rm cross} \sim f_{\rm cross} \sim \lambda^{-\sigma^\prime}$ with
$\sigma^\prime\simeq 3.5$ (see inset in the
Fig. \ref{fig:aval-new-scaled}). 
The role of critical $\lambda$
at which the distribution of forward avalanches follows a power law
is assumed by the value $\lambda=0$. This result is easy to
understand. In fact, in models with fixed (or at least bounded) 
connectivity $c$, the critical $\lambda$ is roughly $1/c$. As will be
shown in the next section, in our
case the size of the system and average connectivity grows without
limits, and thus the critical $\lambda$ tends to zero. Note that it is
 difficult to see this
result without resort to the data collapse
(\ref{eq:collapse}). Indeed, for any finite time of the simulation,
the connectivity and the system size reaches only a limited value and 
the critical $\lambda$   
seen in the distribution of forward avalanches has apparently non-zero
value. 

\subsection{Comparison with the Bak-Sneppen model}
If we compare the above findings with
the BS model, we can deduce that in our model, with $p=1$, the
exponent $\tau$ corresponds to the usual forward-avalanche exponent,
while the exponent $\tau^\prime$ is new. 
The above described scaling (\ref{eq:collapse}) breaks down
for $p<1$ because the connectivity and the system  size are limited
(cf. next section).

The main difference from the usual BS model is the existence of the
second power-law regime, for $s\gg s_{\rm cross}$. It can be
particularly well observed for values of $\lambda$ close to $1$, where
the crossover avalanche size $s_{\rm cross}$ is small. 
We have seen that
such avalanches start and end mostly when number of units is 
close to its minimum value equal to 1. 
Between these events the evolution of the number of units is
essentially a random walk, because singular extinctions are rare
\cite{sla_ko_99}.
This fact
can explain, why the exponent $\tau^\prime$ is not too far from the
value $3/2$ corresponding to the distribution of first returns to the
origin for the random walk. The difference is probably due to the
presence of singular extinctions.

We measured also the distribution of barriers $P(b)$ and the
distribution of barriers on the extremal site 
$P_{\rm min}(b_{\rm min})$.  
In Fig. \ref{fig:bar-distr-new} we can compare the results for
$p=1$ and $p=0.95$. The sharp step observed in BS model is absent
here, because the connectivity is not uniform. (For comparison, we
measured also the barrier distribution in the model of
Ref. \cite{chri_do_ko_sne_98}, where the network is static, but the
connectivity is not uniform. Also in that case the step was absent and the
distribution was qualitatively very similar to the one shown in
Fig. \ref{fig:bar-distr-new}.) The large noise level for $b$ close to
1 is due to the fact that units with larger $b$ undergo mutations rarely.  

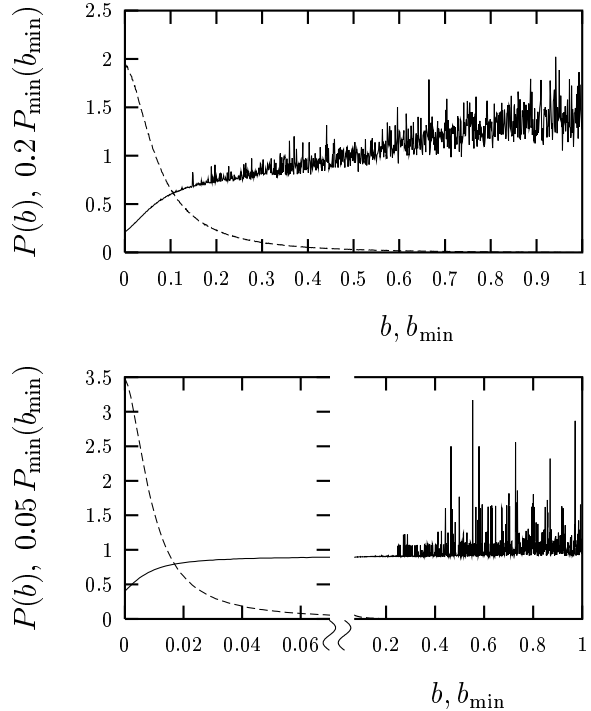
\slaninafigurehh{bar-distr-new}{Distribution of barriers $b$ (full
line) and
minimum barriers $b_{\rm min}$ (dashed line) for $p=0.95$ (upper plot)
and $p=1$ 
(lower plot). In both cases the number of time steps was $10^7$.}

\section{Network geometry}
\label{sec:geometry}
In this section we analyze the geometrical properties of the network 
and their dependence on the parameter $p$.

\subsection{Size of the network}
The first important feature of the networks created by the dynamics of
the model is their size, or the number of units within the
network. This is a strongly fluctuating quantity, but on average it
grows initially and after some time it saturates and keeps fluctuating
around some average value, which depend on $p$. 
Fig. \ref{fig:ns-distr-aver-ns} shows the probability 
distribution of number of units $n_{\rm u}$
for several values of $p$. The average number of units 
$\langle n_{\rm u}\rangle$
was computed from these distributions and its dependence on $p$
is shown in the inset of Fig. \ref{fig:ns-distr-aver-ns}. 
We can see
that the average network size diverges for $p\to 1$ as a power law, 
$\langle n_u\rangle \propto (1-p)^{-\alpha_n}$ with
 the exponent $\alpha_n \simeq 0.8$.

We can see from Fig. 
\ref{fig:ns-c-distr-98-no}
that the distribution of number of units has an exponential tail. This
corresponds to the fact that the time evolution of the network size
is a random walk with reflecting boundary at $n_{\rm u} =1$, with a
bias to lower values, caused by the singular extinctions (for the
analysis of biased random walks repelled from zero see e. g. \cite{so_co_97}).
From the decrease of average size with decreasing $p$ we deduce that
the bias due to the singular extinctions has larger effect for smaller
$p$, i. e. if the new unit created in a speciation event has fewer
links to the neighbors.  

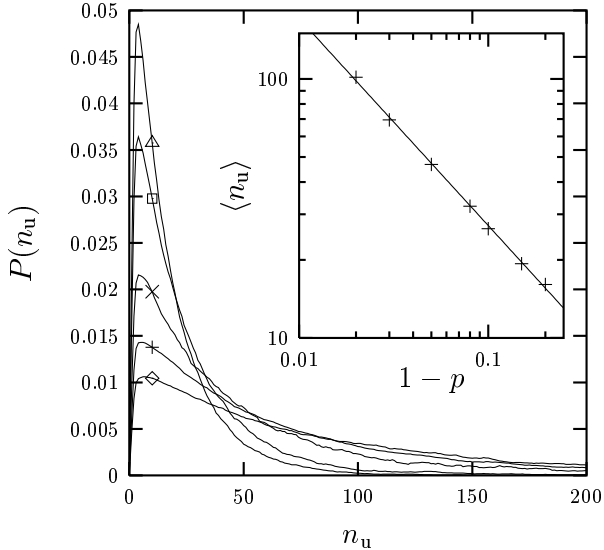
\slaninafigureh{ns-distr-aver-ns}{Distribution of number of units 
for different values of $p$
($\triangle$ 0.85, $\Box$ 0.9, $\times$ 0.95, $+$ 0.97, $\Diamond$ 0.98).
Data are averaged over $10^8$ time steps.
Inset: Dependence of averaged number of units on $p$. The solid line 
corresponds to the power law $\langle n_u\rangle \propto (1-p)^{-0.8}$.
}
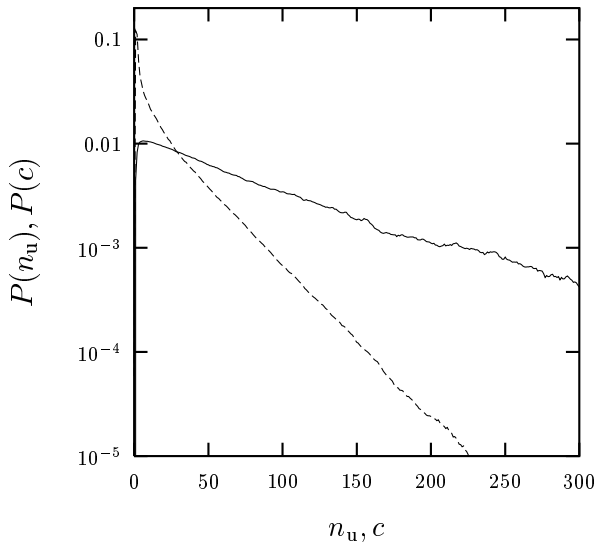
\slaninafigureh{ns-c-distr-98-no}{Distribution of the number of units
 (full line) and connectivity (dashed line), for $p=0.98$, averaged
over  $10^8$  time steps.}

\subsection{Connectivity}
In Fig. \ref{fig:c-distr-98-no}
we show the probability distribution of the connectivity of network
sites $P_{\rm all}(c)$ and distribution of connectivity of the
extremal unit $P_{\rm extremal}(c)$. We can observe the tendency that
the extremal unit has larger connectivity than average. This is in
accord with the findings of Ref. \cite{chri_do_ko_sne_98} obtained on
static networks. It can be also easily understood intuitively. Indeed,
in a mutation event the barriers of neighbors of the mutated unit are changed. 
So, the neighbors have enhanced probability to be extremal in the next
time step. Therefore, the sites with higher number of neighbors have
larger probability that a mutation occurs in their neighborhood and
that they are then mutated in the subsequent step.

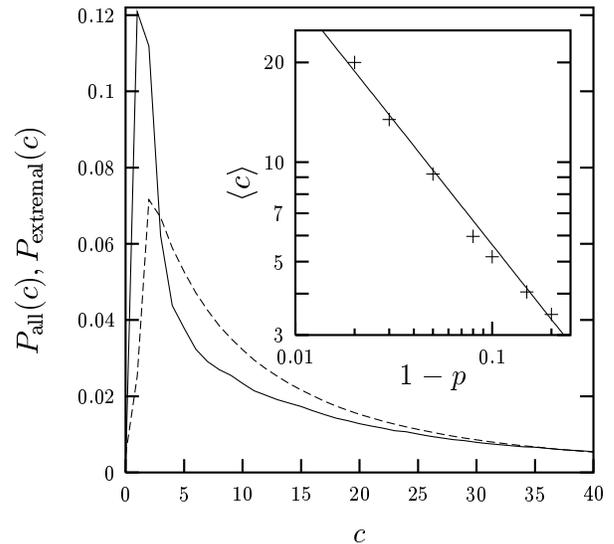
\slaninafigureh{c-distr-98-no}{Distribution of the connectivity for all 
sites (full line) and for extremal sites only (dashed line), 
in the stationary regime for $p=0.98$, averaged over $10^8$ time steps.
Inset: Dependence of averaged connectivity on $p$. The solid line 
corresponds to the power law $\langle c\rangle \propto (1-p)^{-0.75}$.
}

The average connectivity  
$\langle c\rangle$
computed from the distributions $P_{\rm all}(c)$
is shown in the inset of Fig. \ref{fig:c-distr-98-no}. 
We can observe that analogically to the system size also the average
connectivity diverges for $p\to 1$ as a power law, but the value of the
exponent is slightly different. We find 
$\langle c \rangle \propto (1-p)^{-\alpha_c}$ with
 the exponent $\alpha_c \simeq 0.75$. From the data available we were
not able do decide, whether the exponents $\alpha_n$ and $\alpha_c$
are equal within the statistical noise.

In Fig. \ref{fig:ns-c-distr-98-no}
we can see that also the distribution of connectivity has an
exponential tail, similarly to the distribution of network size. 
We measured also the joint probability density $P(n_{\rm u},c)$ for
the number of units 
and the connectivity. The result is shown as a contour plot in
Fig. \ref{fig:c-n-distr}.
We can see that also for large networks (large $n_{\rm u}$) the most
probable connectivity is small and nearly independent on $n_{\rm
u}$. This means that the overall look of the network created by the
dynamics of our model is that there are a few sites with large
connectivity, surrounded by many sites with low connectivity.

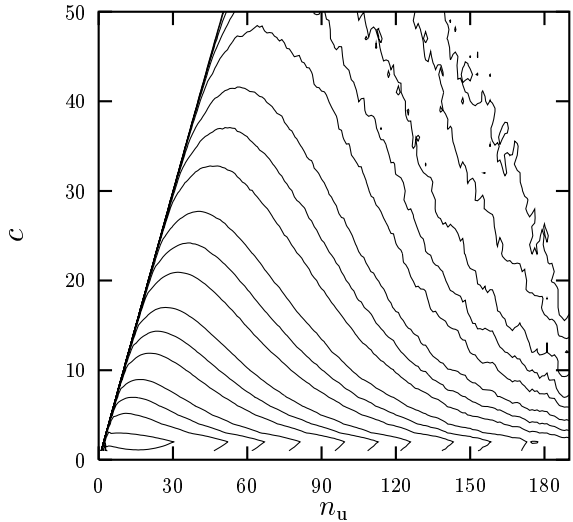
\slaninafigureh{c-n-distr}{Contour plot of the joint probability
density $P(n_{\rm u},c)$ for number of units and connectivity,
for $p=0.8$, averaged over  $3\cdot 10^9$ time steps. 
The contours correspond to the following values of the
probability density (from inside to outside): 
$5\cdot 10^{-R}$, $2\cdot 10^{-R}$,$10^{-R}$, with orders $R=3,4,5,6,7,8$.}
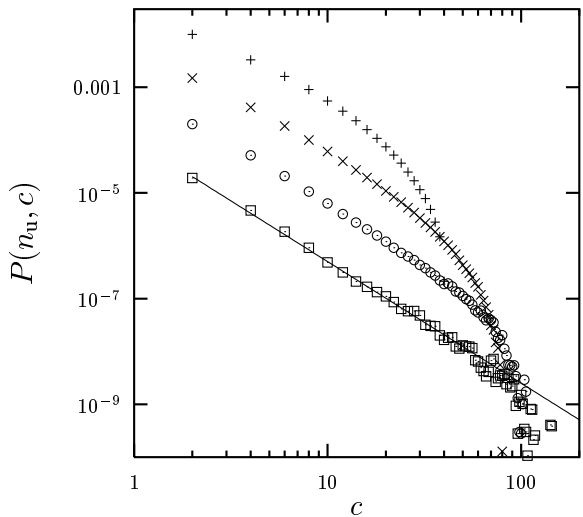
\slaninafigureh{c-distr-8-loglog}{Distribution of connectivity for
fixed number of units, for $p=0.8$ and sizes $n_{\rm u}=40$ ($+$), 80
($\times$), 120 
($\odot$), 170 ($\Box$). The straight line is a power-law with
exponent $-2.3$. The data are the same as those used in
Fig. \ref{fig:c-n-distr}.}

An interesting observation can be drawn from the results shown in
Fig. \ref{fig:c-distr-8-loglog}. It depicts the joint probability
density as a function of connectivity at fixed system sizes. We can
see that for smaller system 
sizes, closer to the average number of units, the distribution is
exponential, while if we increase the system size a power-law
dependence develops. For example for the system size fixed at $n_{\rm
u}=170$ we observe a power-law behavior $P(n_{\rm u},c)\sim c^{-\eta}$
nearly up to the geometric cutoff, $c< n_{\rm u}$. The value of the
exponent was about $\eta\simeq 2.3$. 

This finding may be in accord
with the power-law distribution in growing networks
\cite{bar_alb_99,dor_men_sam_00}. Indeed, in our model the power-law
behavior applies only for networks significantly larger than the
average size. Such networks are created during time-to-time
fluctuations leading to temporary expansion of the network. So, the
power-law is the trace of expansion periods in the network evolution,
corresponding to continuous growth in the model of \cite{bar_alb_99}.
The preferential attachment, which is the second key ingredient in
\cite{bar_alb_99} has also an analog in our model; highly connected
units are more likely to be mutated, as was already mentioned in the
discussion of Fig.  \ref{fig:c-distr-98-no}. However, here the
preference of highly connected sites is a dynamical phenomenon,
resulting from the extremal dynamics rules of our model.

\subsection{Clusters}
As noted already in the Sec. \ref{sec:model}, the network can be split
into several disconnected clusters.  The clusters cannot merge, but
they may vanish due to extinctions. We observed qualitatively that
after initial growth the number of clusters exhibits stationary
fluctuations around an average value, which increases when $p$ approaches
1. We measured both the distribution of the number of 
clusters and the distribution of their sizes. In
Fig. \ref{fig:nucl-distr-98-loglog} 
we show the distribution of the number of clusters. The most probable
situation is that there is only a single cluster. However, there is a
broad tail, which means that even large number of clusters can be
sometimes created. The tail has a power-law part with an exponential
cutoff. The value of the exponent in the power-law regime $P(n_c)\sim
n_c^{-\rho}$ was about $\rho\simeq 1.2$. We have observed that the
width of the power-law regime is larger for larger $p$. This leads us
to the conjecture that in the limit $p\to 1$ the number of clusters is
power-law distributed.

On the other hand, the distribution of cluster sizes shown in  
Fig. \ref{fig:sc-distr-98-no}
has maximum at very small values. This is due to two effects. First,
already the distribution of network size has maximum at small sizes,
and second, if the network is split into many clusters, they have
small size and remain unchanged for long time. The reason why small
clusters change very rarely (and therefore can neither grow nor
disappear) can be also seen from
Fig. \ref{fig:sc-distr-98-no}, where the distribution of sizes of the
clusters containing the extremal site is shown. 
The latter
distribution is significantly different from the size distribution for
all clusters and shows that the extremal site belongs mostly to large
clusters. In fact, we measured also the fraction indicating  how often
the extremal unit is in the largest cluster,  if there are more than
one cluster. 
For the same run from which the data  shown in
Fig. \ref{fig:sc-distr-98-no} were collected, we found that this
fraction is 0.97, i. e. very close to 1.
A similar ``screening effect'' was reported also in the Cayley tree models
\cite{va_au_95}: the small isolated portions of the network are
very stable and nearly untouched by the evolution. 

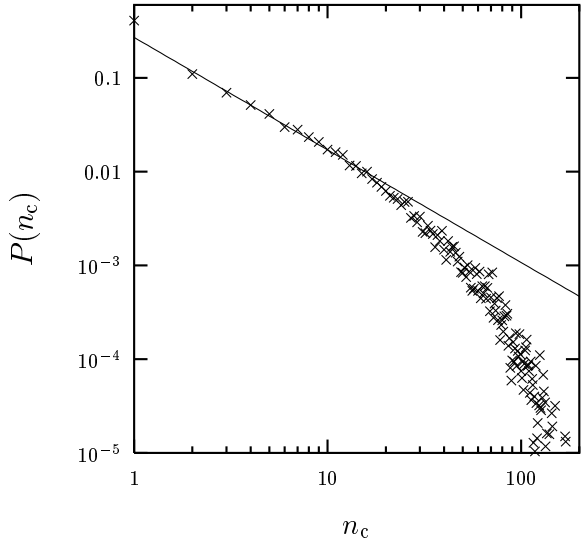
\slaninafigureh{nucl-distr-98-loglog}{Distribution of the number 
of clusters, for $p=0.98$. The straight line is a power law with
exponent $-1.2$. The data were averaged over 3 independent
runs, $5\cdot 10^8$, $5\cdot 10^8$, and $10^8$ time steps long.}
\slaninafigureh{sc-distr-98-no}{
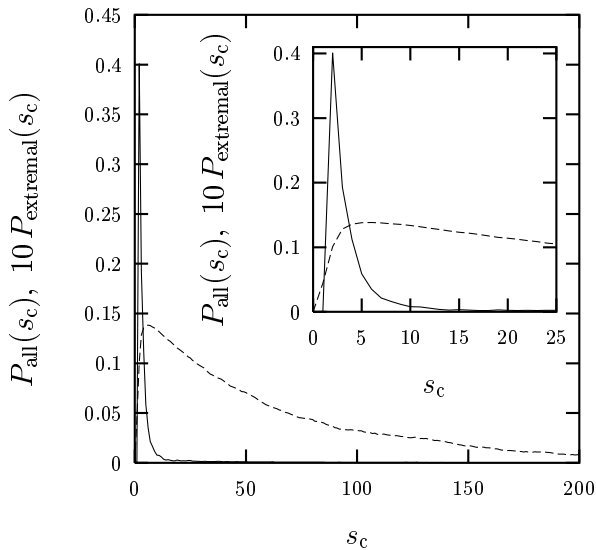
Distribution of the cluster sizes $s_c$
for $p=0.98$, averaged over $10^8$ time steps.
Full line - all clusters, dashed line - clusters containing the extremal
site. Inset: Detail of the same distribution.
}

\subsection{Mean distance}
An important feature of a random network is also the mean distance $\bar{d}$
between two sites, measured as minimum number of links, which should be passed
in order to get from one site to the other. In $D$-dimensional
lattices, the mean distance depends on the number of sites $N$ as
$\bar{d}\sim N^{1/D}$, while in completely random networks the dependence is
$\bar{d}\sim \log N$. In the small-world networks, the crossover from the
former to the latter behavior is observed \cite{wa_stro_98,newman_00}.

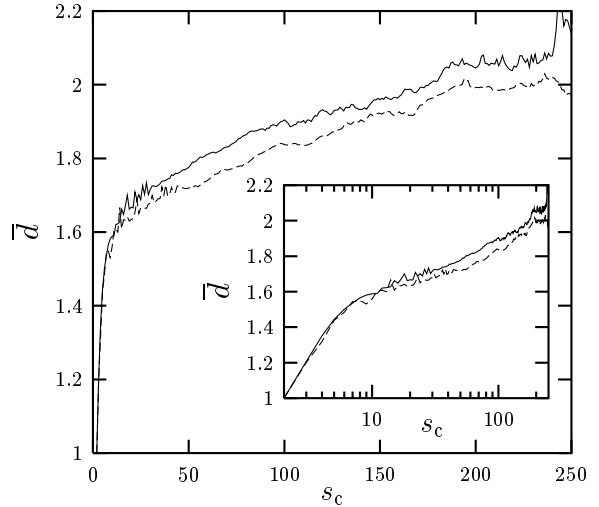
\slaninafigureh{distance}{Dependence of the average distance of two sites 
within the same cluster on the cluster size, for $p=0.95$ (full line) and 
$p=0.97$ (dashed line), averaged over for $10^7$ time steps.
 In the inset we show the same data in the log-linear scale.}

The dependence of the average distance within a cluster on the size of
the cluster in our model is shown in
Fig. \ref{fig:distance}. We can observe global tendency to decrease
$\bar{d}$ when increasing $p$. This result is natural, because larger
$p$ means more links from a randomly chosen site and thus shorter
distance to other sites. The functional form of the size dependence is not
completely clear. However, for larger cluster sizes, greater than
about 25, the dependence seems to be faster than logarithmic,
as can be seen from the inset in Fig. \ref{fig:distance}. So, the
networks created in our model seem to be qualitatively different from
the random networks studied previously, as far as we know.

\section{Conclusions}
\label{sec:conclusions}

We studied an extremal dynamics model
motivated by biological evolution on dynamically evolving 
random network. The properties of
the model can be tuned by the parameter $p$, the probability that a
link is inherited in the process of speciation. For $p=1$ the model is
self-organized critical and the average system size and connectivity
grows without limits. Contrary to the usual BS model, we find two
power-law regimes with different exponents in the statistics of
forward $\lambda$-avalanches. The crossover avalanche
size depends on $\lambda$ and diverges for $\lambda\to 0$ as a power
law. The reason why the critical $\lambda$ is zero in this model is
connected with the fact that time-averaged connectivity diverges for
$p=1$.

We investigated the geometrical properties of the random networks for
different values of $p$. The average network size and average
connectivity diverge as a power of $1-p$. The probability distribution
of system sizes has an exponential tail, which suggests that the
dynamics of the system size is essentially a biased random walk with a
reflecting boundary, The value of the bias grows with decreasing $p$. 
The joint distribution of size and connectivity shows that even for
large network sizes the most probable connectivity is low. Hence,
there are few highly-connected sites linked to the majority of sites
with small connectivity.
Moreover,
the situations where the system size is far above its mean value are
characterized by power-law distribution of connectivity, like in the
models of growing networks with preferential attachment. 

The network can consist of several mutually disconnected
clusters. Even though the most probable situation contains only a
single cluster, the distribution of cluster numbers has a broad
tail, which shows a power-law regime with exponential cutoff. We
observed the ``screening effect'', characterized by very 
small probability that the extremal site is found in any other cluster
than the largest one. So, there is a central large cluster, where
nearly everything happens, surrounded by some small peripheral
clusters, frozen for the major part of the evolution time.

We measured also the mean distance measured along the links within one
cluster. The distance grows very slowly with the cluster size;
however, the increase seems to be faster than logarithmic.

Summarizing, we demonstrated that the extremal dynamics, widely used
in previous studies on macroevolution in fixed-size systems is useful 
in creating random networks of variable size.
It would be of interest to compare the properties of the networks
created in our model with food webs and other networks found in 
the nature. For example the studies of food webs in isolated ecologies
\cite{wil_mar_00} give for network sizes about 30
average connectivities in the range 
from 2.2 to 9, which is not in contradiction with 
the findings of our model. However, more precise comparisons are
necessary for any reliable conclusions about real ecosystems.

\pagebreak

\section*{Acknowledgments}

We wish to thank K. Sneppen, A. Marko\v{s} and A. P\c{e}kalski for
useful discussions.  

%\pagebreak

\pagebreak

\end{document}